# Heavy-Ion Physics with ALICE

Federico Antinori[a,b] (for the ALICE[*] Collaboration)

[a] Istituto Nazionale di Fisica Nucleare, Padova, Italy
[b] CERN, Geneva, Switzerland

**Abstract**

The ALICE detector, expected to start operating at the Large Hadron Collider this year, was designed specifically for the study of heavy-ion collisions. In this paper we recall the main features of the apparatus and give some examples of the expected physics performance.

1. **Introduction**

The Large Hadron Collider (LHC), scheduled to start operating in late 2007, will allow to study collisions of heavy ions in a new, hitherto unexplored energy regime. The ALICE detector, presently in the final phases of construction, has been specifically designed for the harsh environment of central Pb-Pb collisions at the LHC, where thousands of charged particles per unit rapidity are expected.
In this paper, after recalling some of the key parameters of the LHC, we shall give an overview of the ALICE experimental apparatus; we shall then discuss some specific examples of the expected physics capabilities and comment on the startup plans.

2. **LHC as an ultrarelativistic heavy-ion collider**

The LHC, designed to collide protons at a c.m.s. energy $\sqrt{s} = 14$ TeV, will also accelerate ions up to the same magnetic rigidity and allow the study of both symmetric systems (e.g. Pb-Pb) and asymmetric collisions, such as proton-nucleus.
In table 1, we give examples of the c.m.s. energy and expected luminosity in the LHC for some collision systems. The typical expected yearly running times are of the order of $10^7$ s for proton-proton collisions and $10^6$ s for the heavier systems.

---

[*] for the full author list see appendix

| Collision system | $\sqrt{s_{NN}}$ (TeV) | $L_0$ (cm$^{-2}$s$^{-1}$) | $\sigma_{geom}$ (b) |
|---|---|---|---|
| pp | 14.0 | $10^{34}$ * | 0.07 |
| PbPb | 5.5 | $10^{27}$ | 7.7 |
| pPb | 8.8 | $10^{29}$ | 1.9 |
| ArAr | 6.3 | $10^{29}$ | 2.7 |

Table 1. Examples of c.m.s. energy and luminosities expected for different collision systems in the LHC. The geometrical cross sections are also given.
* due to the limited rate capability of the ALICE detector, we plan to limit the luminosity in our interaction region to a maximum of $10^{31}$ cm$^{-2}$s$^{-1}$

## 3. The ALICE experiment

The ALICE experimental apparatus, sketched in figure 1, consists of two main components: a central detector and a muon arm. The central detector, covering mid-rapidity ($|\eta| < 0.9$) over full azimuth, is contained inside a large solenoidal magnet providing a magnetic field $|B| < 0.5$ T. It includes, from the interaction region outwards, a vertex detector – the Inner Tracking System (ITS) – made of six layers of silicon microdetectors (Si pixels, Si drifts and Si strips), a large Time Projection Chamber (TPC) (see [1]), a Transition Radiation Detector (TRD) for electron identification and a Time Of Flight detector (TOF) (see [2]) for the identification of pions, kaons and protons. The central detector is completed by two detectors with reduced azimuthal acceptance: the High Momentum Particle Identification Detector (HMPID) – an array of ring imaging Cherenkov detectors for particle identification at high momentum – and the PHOton Spectrometer (PHOS) – a high resolution electromagnetic calorimeter made of high density scintillator crystals (see [3]).
The muon arm, covering the pseudorapidity interval $-4.0 < \eta < -2.4$, consists of a hadron absorber positioned very close to the interaction region, a spectrometer made of a dipole magnet with a field integral of 3 Tm and five tracking stations, a second iron filter and two trigger stations. Besides the central detector and the muon arm, the ALICE apparatus also includes a set of forward detectors: the Photon Multiplicity Detector (PMD), the Forward Multiplicity Detector (FMD) – extending the coverage for the measurement of charged particle multiplicity to the range $-3.4 < \eta < 5.1$ – the Zero Degree Calorimeters (ZDC) and a system of trigger scintillators and quartz counters (T0 and V0). The apparatus is completed by an array of scintillators for triggering on cosmic rays (ACORDE). A detailed description of the ALICE apparatus can be found in [4].
The latest addition to the ALICE design is an ElectroMagnetic CALorimeter (EMCAL) covering the pseudorapidity interval $|\eta| < 0.7$ and 110º in azimuth [5].

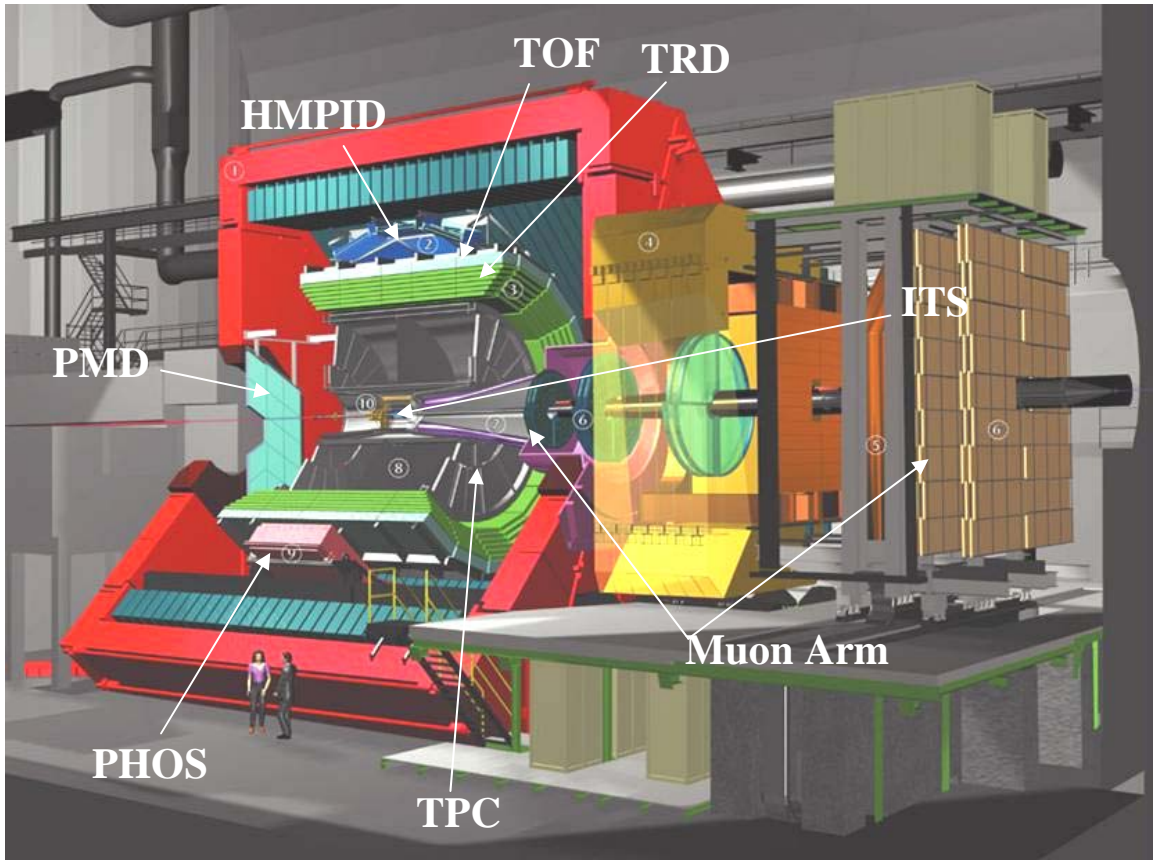

Figure 1: Sketch of the ALICE layout

## 4. Physics performance: examples

The ALICE detector is designed to address and correlate a wide palette of physics observables. A detailed discussion of the expected physics performance can be found in [6]. In this paper, we shall briefly illustrate three specific examples: the measurement of heavy flavour production and of jet production in the central detector and the measurement of the production of quarkonia in the muon arm.

### 4.1 Heavy flavours in the central detector

Suppression of the production of high transverse momentum ($p_T$) particles ("quenching") was observed at the Relativistic Heavy-Ion Collider (RHIC) and interpreted as due to energy loss at the partonic level (see for instance [7-10] and references therein). Theoretically [11], the energy loss effect is expected to be parton-specific (stronger for gluons than for quarks due to the former's higher colour charge) and flavour-specific (weaker for heavy than for light quarks, due to the so-called dead-cone effect [12]). One would therefore expect less high $p_T$ quenching for heavy flavour particles, that originate from heavy quark jets, than for light flavour particles, that are created in both gluon and (mostly light) quark jets.

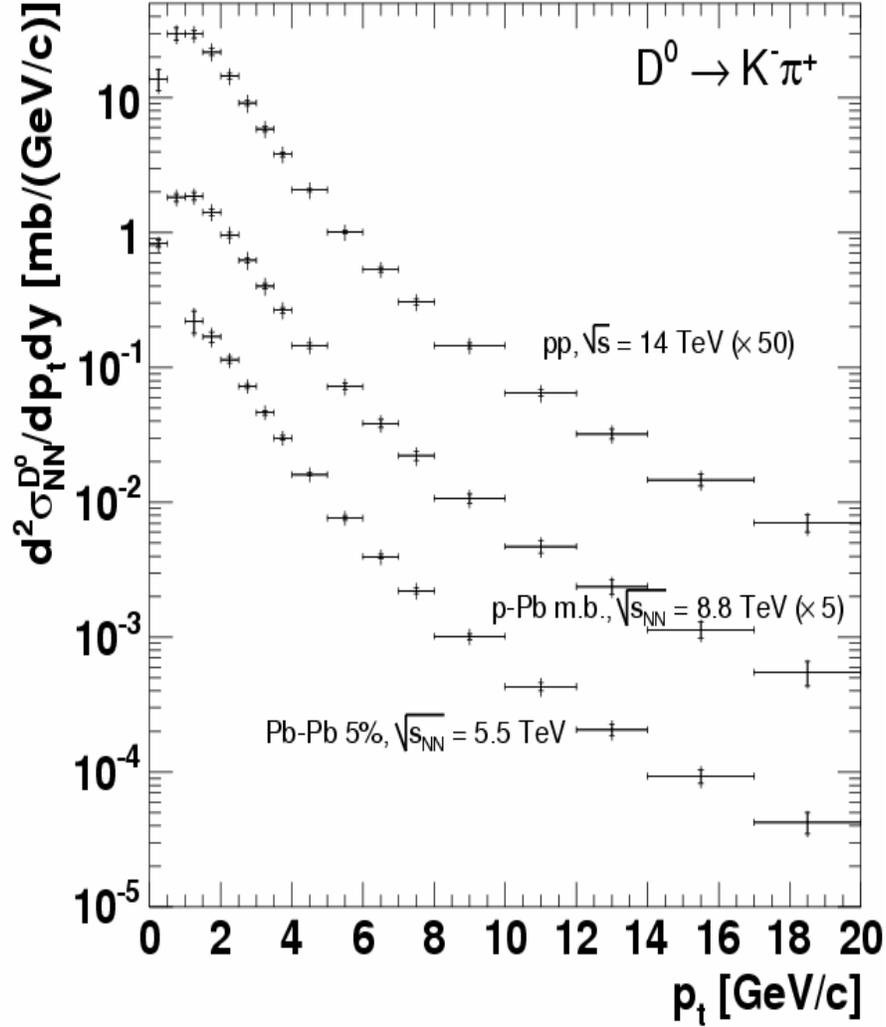

Figure 2: Expected performance for the measurement of the $D^0$ $p_T$-differential cross section for a statistics corresponding to one yearly run for pp ($10^7$ events), p-Pb ($10^6$ events) and Pb-Pb ($10^6$ events)

Experimentally, however, at RHIC the quenching of "non-photonic electrons", thought to be mostly generated in heavy flavour decays, is found to be rather similar to that of light hadrons [13, 14]. The interpretation of these results is complicated by the fact that, while the quenching effect is expected to be very different for charm and beauty quarks, there is no obvious way experimentally to disentangle the charm and beauty contributions to the RHIC non-photonic electron samples.

At the LHC, where much higher heavy flavour yields are expected (over 100 $c\bar{c}$ pairs and about 5 $b\bar{b}$ pairs per central Pb-Pb collision) one should be able to study the production of heavy flavour hadrons in heavy-ion collisions at an unprecedented level of detail (see also [15]). In addition, unlike the RHIC detectors, ALICE is equipped with a vertex detector (the ITS) which allows to measure the separation from the primary vertex of the heavy flavour decay tracks, with a track impact parameter resolution expected to be better than 50 μm for $p_T$ > 1.3 GeV/c. Therefore, we anticipate being able to fully reconstruct heavy flavour decays and to measure separately the production of charm and beauty hadrons. As an example, in figure 2 we show the expected performance for data samples corresponding to the statistics expected for one yearly run for the measurement of the production of the $D^0$ via reconstruction of its decay in the $K^-\pi^+$ channel, using information from the ITS, TPC and TOF detectors.

In ALICE, good electron identification capabilities can be achieved combining the information from the TRD with the energy loss of the candidate track in the TPC. Beauty production can then be measured by selecting a sample of sufficiently high $p_T$ electrons well separated from the primary vertex. To give an example, by selecting electrons with $p_T > 2$ GeV/c and missing the primary vertex by more than 200 μm, we expect a high statistics ($\sim 8 \cdot 10^4$ $e^\pm$) sample with a beauty purity of about 80%. The residual contamination is essentially due to charm electrons, and can be subtracted with good precision once the charm yield is measured via the reconstruction of D decays [6]. The comparison of the quenching for beauty and charm hadrons will then allow to isolate the parton mass dependence of the effect, while by comparing the D and light hadron results one should be able to extract information on its colour charge dependence.

**4.2 Jets**

At the LHC, jets will be abundantly produced. We expect around one jet with energy above 20 GeV per central Pb-Pb event and about $10^5$ jets with energy above 200 GeV in $10^6$ s of Pb-Pb running. Such high energy jets will be easily visible and reconstructable even in the very high multiplicity environment of central Pb-Pb collisions at the LHC (the example for the case of a 100 GeV jet embedded in a central Pb-Pb event – with a charged particle multiplicity density $dN_{ch}/dy = 6000$ – is shown in figure 3).

The addition of the EMCAL to the ALICE layout will substantially enhance the experiment's performance for jet physics, by improving the precision of the measurement of the jet energy and by introducing the capability of triggering on high energy jets, thereby extending the kinematic reach.

The reconstruction of jets in heavy-ion collisions presents some difficulties due to the very high background, in particular dictating the choice of the jet cone size $R = \sqrt{(\Delta\eta)^2 + (\Delta\varphi)^2}$, which has to be reduced with respect to what is normally used in elementary collisions ($R \sim 1$). The introduction of a $p_T$ cut should also help reduce the soft background.

Parton energy loss is expected to soften the in-medium fragmentation function with respect to the vacuum case, depleting the abundance of hadrons at high $z$ ($= p_T^{hadron}/E_T^{jet}$) and enhancing it at low $z$. The measurement of the modification of the jet fragmentation function in central heavy-ion collisions – relative for instance to peripheral collisions – is therefore expected to provide information about the quenching mechanism. In figure 4 we show the result of a quenching weight [16] calculation [6] of the effect of parton energy loss due to gluon radiation on the fragmentation function for jets of around 125 GeV (a medium transport coefficient $\hat{q} = 50$ GeV$^2$/fm was used). The expected ALICE performance for the measurement of the modification of the fragmentation in a $10^6$ s Pb-Pb run is also shown.

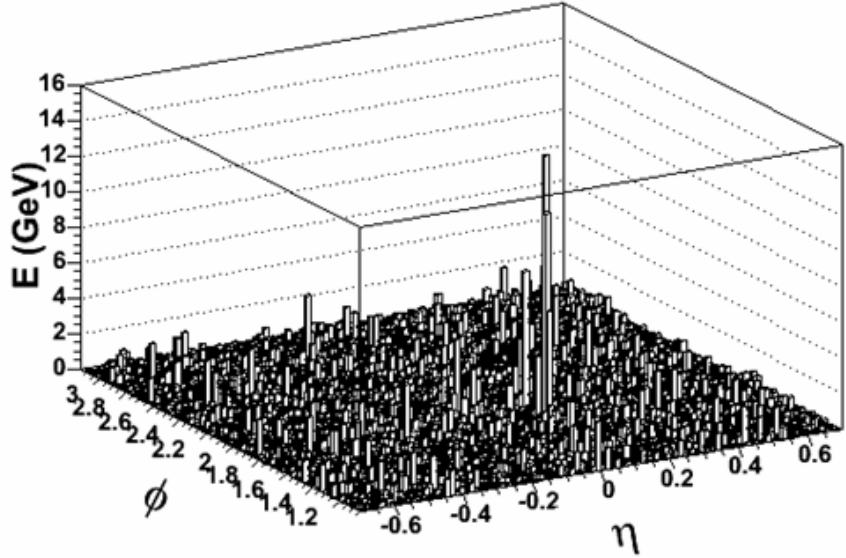

Figure 3: A 100 GeV jet embedded in a simulated central Pb-Pb event with $dN_{ch}/dy = 6000$

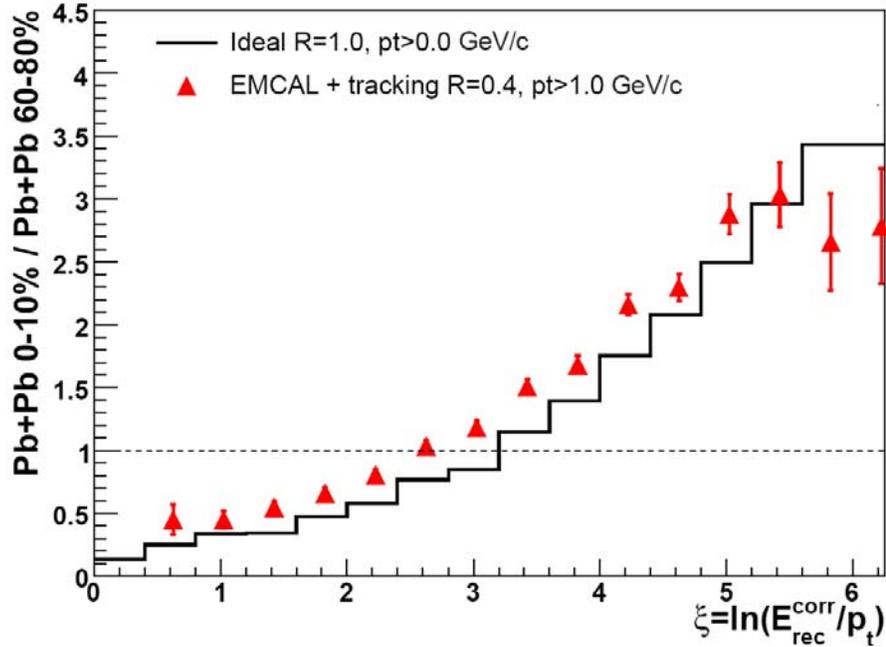

Figure 4: Expected ALICE performance (triangles) for the measurement of the ratio of the fragmentation functions for central (modified) and peripheral (taken to be unmodified) collisions for jets with a mean energy around 125 GeV, as a function of $\xi = \ln(1/z)$. The ideal case of perfect jet reconstruction is shown as a histogram. The modification to the fragmentation function was calculated using quenching weights [16] with a medium transport coefficient $\hat{q} = 50$ GeV$^2$/fm

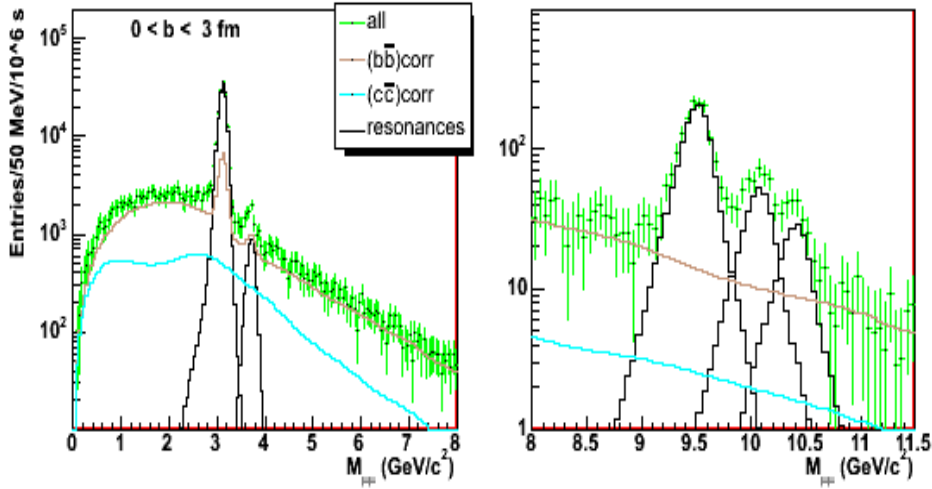

Figure 5: Expected ALICE performance (after subtraction of the combinatorial background) for the measurement of the dimuon spectrum in the muon arm for a $10^6$ s Pb-Pb run: charmonium (left) and bottomonium (right) mass ranges

**4.3 Quarkonia in the muon arm**

Quarkonia production is one of the prime observables in ultrarelativistic heavy-ion physics, following an early prediction [17] that the production of quarkonia would be suppressed in the Quark-Gluon Plasma (QGP), since the strong interaction equivalent of Debye screening would hinder the binding of the heavy $Q\bar{Q}$ pair in the deconfined medium. Anomalous suppression of the production of J/ψ was indeed observed in central Pb-Pb collisions at the CERN Super Proton Synchrotron (SPS), on top of the expected "nuclear" suppression systematics observed with lighter systems [18]. As an alternative to the QGP explanation, it was also proposed that the anomalous suppression could be due to an effect of dissociation of the produced J/ψ by interaction with comoving hadrons. In both cases, it was expected that the effect would be stronger at RHIC [19-21]. Experimentally, however, the amount of J/ψ suppression observed at RHIC turned out to be very similar to the SPS one [22]. Better agreement with the data is obtained if some mechanism of J/ψ regeneration [23, 24], e.g. by recombination, is introduced [19, 20, 25-27]. In such a case, the similarity of the suppression at SPS and RHIC would be the result of a somewhat fortuitous cancellation of the extra suppression at RHIC by an increase in the $c\bar{c}$ abundance. At the LHC, however, due to the much higher $c\bar{c}$ cross section, regeneration should then have the upper hand. The suppression would then be reduced, and may even turn into an enhancement [28].

In ALICE, $\mu^+\mu^-$ pairs from the decay of quarkonium particles should generate a very clean signal in the muon arm (see also [15]). An important experimental issue: at the LHC energy one expects a significant (20 – 30 %) contribution to the J/ψ yield from B decays. This contribution will be controlled by measuring open B production both in the central detector, as discussed above, and directly in the muon arm [6].

The simulated performance on the dimuon mass spectrum after subtraction of the combinatorial background for $10^6$ s Pb-Pb running is shown in figure 5. We expect a mass resolution around 70 MeV (100 MeV) for charmonia (bottomonia) in central Pb-Pb collisions. The detector acceptance extends down to essentially zero $p_T$ for both charmonia and bottomonia. The statistics collected in a $10^6$ s Pb-Pb run should allow to measure out to a $p_T$ around 20 GeV for J/ψ, and around 10 GeV for ϒ(1S) and ϒ(2S).

## 5. Startup plans

The installation of the various ALICE components (detectors, infrastructure, services) is proceeding according to schedule. We plan to close the experiment and start commissioning in August 2007. A first proton-proton commissioning run at $\sqrt{s} = 0.9$ TeV is foreseen for November-December 2007. The first full energy proton-proton run at $\sqrt{s} = 14$ TeV is then planned for 2008. These runs will not only provide the baseline, reference data for the ALICE heavy-ion program, but will also be analyzed within the dedicated proton-proton physics program (not discussed in this contribution, see [29] at this conference) that exploits the very good expected proton-proton performance of the experiment, notably in the low-$p_T$ sector.
An initial low luminosity Pb-Pb run is expected to follow the first long proton run.
The ITS, TPC, HMPID, muon arm, PMD, V0, T0, ZDC and ACORDE detector are all expected to be completely installed by the summer, while partial installation only is planned for the PHOS (1/5), TOF (9/18), TRD (3/18) and Data Acquisition System (30%).

## 6. Conclusions

After about 15 years of preparation, the ALICE detector is finally becoming a reality. We are looking forward to a very exciting period. The experiment was designed based on known signals and observables, and we have discussed above a few examples of the expected physics capabilities. Most importantly however, ALICE is a very complete and versatile instrument – dedicated and optimized for heavy ion physics – to explore the unchartered LHC territory.